\documentclass[twocolumn,superscriptaddress,aps, prl]{revtex4}
\usepackage[intlimits]{amsmath}
\usepackage{amsfonts}
\usepackage{graphics}
\usepackage{subfigure}
\usepackage[usenames]{color}
\usepackage{epstopdf}
\usepackage{color}
\usepackage{indentfirst}
\usepackage{graphicx}
\usepackage{amsmath}
\usepackage{amsfonts}

\DeclareMathSizes{10.5}{10.5}{1}{1}

\usepackage{bm}

\newcommand{\ket}[1]{\left\vert#1\right\rangle}
\newcommand{\bra}[1]{\left\langle#1\right\vert}

\newcommand{\average}[1]{\left\langle#1\right\rangle}

\begin{document}

\title{Nonequilibrium critical scaling from quantum thermodynamics}

\author{Abolfazl Bayat}
\affiliation{Department of Physics and Astronomy, University College London, Gower Street, LondonWC1E 6BT, United Kingdom}

\author{Tony J. G. Apollaro}
\affiliation{NEST, Scuola Normale Superiore \& Istituto di Nanoscienze-CNR, I-56126 Pisa, Italy}
\affiliation{Dipartimento di Fisica e Chimica, Universit\`{a} degli Studi di Palermo, Via Archirafi 36, I-90123 Palermo, Italy}
\affiliation{Centre for Theoretical Atomic, Molecular, and Optical Physics, School of Mathematics and Physics, Queen's University Belfast,BT7,1NN, United Kingdom}
\affiliation{International Institute of Physics, Universidade Federal
do Rio Grande do Norte, 59078-400 Natal-RN, Brazil}

\author{Simone Paganelli}
\affiliation{International Institute of Physics, Universidade Federal
do Rio Grande do Norte, 59078-400 Natal-RN, Brazil}
\affiliation{Dipartimento di Scienze Fisiche e Chimiche, Universit\`{a} dell'Aquila, via Vetoio, I-67010 Coppito-L'Aquila, Italy}

\author{Gabriele~De~Chiara}
\affiliation{Centre for Theoretical Atomic, Molecular, and Optical Physics, School of Mathematics and Physics, Queen's University Belfast,BT7,1NN, United Kingdom}

\author{Henrik Johannesson}
\affiliation{Department of Physics, University of Gothenburg, SE 412 96 Gothenburg, Sweden}
\affiliation{Beijing Computational Science Research Center, Beijing 100094, China}

\author{Sougato Bose}
\affiliation{Department of Physics and Astronomy, University College London, Gower Street, LondonWC1E 6BT, United Kingdom}

\author{Pasquale Sodano}
\affiliation{International Institute of Physics, Universidade Federal
do Rio Grande do Norte, 59078-400 Natal-RN, Brazil}
\affiliation{Departemento de Fis\'ica Teorica e Experimental,
Universidade Federal do Rio Grande do Norte, 59072-970 Natal-RN, Brazil}
\affiliation{INFN, Sezione di Perugia, Via A. Pascoli, I-06123, Perugia,
Italy}


\begin{abstract}
\noindent

The emerging field of quantum thermodynamics is contributing important results and insights into archetypal many-body problems, including quantum phase transitions. Still, the question whether out-of-equilibrium quantities, such as fluctuations of work, exhibit critical scaling after a sudden quench in a closed system has remained elusive. Here, we take a novel approach to the problem by studying a quench across an impurity quantum critical point. By performing density matrix renormalization group computations on the two-impurity Kondo model, we are able to establish that the irreversible work produced in a quench exhibits finite-size scaling at quantum criticality. This scaling faithfully predicts the equilibrium critical exponents for the crossover length and the order parameter of the model, and, moreover, implies a new exponent for the rescaled irreversible work. By connecting the irreversible work to the two-impurity spin correlation function, our findings can be tested experimentally.

\end{abstract}


\maketitle

\emph{Introduction.--} Out-of-equilibrium thermodynamics of closed many-body systems subject to a
variation of  a Hamiltonian parameter is receiving considerable attention, both experimentally and theoretically \cite{PolkolnikovRMP11}.
The increasing level of control over few-particle quantum systems has allowed to demonstrate experimentally the information-to-energy conversion and the Jarzynski equality \cite{Jarzynski07,TayabeetalNat10, PekolaPNAS14,AnNat15}, and has also inspired various proposals for constructing
quantum engines~\cite{AbahLutzetalPRL12, FialkoPRL12}.
Moreover,  studies of the interplay between quantum thermodynamics, many-body physics, and quantum information, have shed light on fundamental aspects of
thermalisation of closed quantum systems \cite{Eisert2015}, fluctuation-dissipation relations~\cite{CampisiHTRMP11}, and prospects for quantum simulations \cite{GooldHRRS15}.

A central issue is how the presence of a quantum phase transition (QPT) shows up in the out-of-equilibrium thermodynamics after a sudden quench of a Hamiltonian parameter
\cite{PolkolnikovRMP11,Eisert2015,Talkner07,silva08,dorner12,Fuscoetal14,PlastinaetalPRL14,mascarenhas14,PlastinaetalNJP14,ApollaroetalPhysS}. It is now established \cite{silva08} that a second-order QPT is signaled by a discontinuity in the derivative of the {\em irreversible entropy production} \cite{CampisiHTRMP11}, as well as of the {\em variance of the work} \cite{CampisiHTRMP11} (with the derivative taken with respect to the QPT driving  parameter which is being quenched). This is to be contrasted with a first-order QPT, where the derivative of the {\em average work}
exhibits a discontinuity at the transition \cite{mascarenhas14} (with a peak in the irreversible entropy production when the QPT is induced by a local quench ~\cite{ApollaroetalPhysS}). The obvious parallels to the diverging behavior of response functions at a second-order equilibrium QPT prompts the question whether out-of-equilibrium quantities, like the {\em irreversible work} \cite{CampisiHTRMP11}  (which is a measure of the nonadiabaticity of a quantum quench), may also exhibit scaling at criticality. Here, via a novel inroad $-$ studying the quantum thermodynamics for a sudden quench across an impurity quantum critical point $-$ we are able to provide an affirmative answer.

To set the stage, let us recall that while for an ordinary bulk QPT the behavior of thermodynamic quantities after a sudden quench reflects
the discontinuity of a corresponding equilibrium average value of a global observable \cite{dorner12}, the same is not so obvious in an impurity quantum phase transition iQPT~\cite{Vojta06}. This kind of transition does not easily fit into the Ehrenfest-Landau scheme, and the definition of a suitable order parameter is far from a trivial task~\cite{bayat-Schmidt-14,Wang15}. The bulk properties are independent of the impurity, and the critical point is characterized by singularities in the impurity non-extensive contribution to the ground state energy. Hence, one may ask whether, after a local quench of the impurity coupling, the behaviour of nonequilibrium quantum thermodynamic variables can reveal the iQPT?

We shall address this question in the specific case of the two-impurity Kondo model (TIKM) ~\cite{Jayaprakash81}, one of the best studied models supporting an iQPT~\cite{jones1988,jones1989,affleck1992,affleck1995,sire1993,gan1995,zarand2006,sela2011,Mitchell-sela2012,mitchell2012,he2015}. Here, two spin-1/2 quantum impurities are coupled to each other by a Ruderman-Kittel-Kasuya-Yosida (RKKY) interaction, and, in the simplest variant of the theory  \cite{zarand2006}, to separate bulk reservoirs of conduction electrons by Kondo interactions.  When the RKKY interaction dominates, the two impurities form a local spin-singlet state (RKKY phase), while in the opposite limit each of the impurities form a spatially extended singlet state with the electrons in the reservoir to which it is coupled (Kondo-screened phase).
We shall show that the iQPT between these two phases is signaled both by the irreversible work production and the variance of work following a sudden quench.
Remarkably,
the irreversible work shows manifest scaling with well-defined critical exponents, related to known equilibrium critical exponents by scaling laws. Moreover, by means of a small quench approximation for the irreversible work production, we are able to link the latter to the two-impurity spin correlation function, which is amenable to experimental determination. While our findings have broad ramifications, the particulars may be especially timely considering recent breakthroughs in designing and performing measurements on tunable nanoscale realizations of the TIKM~\cite{bork2011,chorley2012,spinelli2014}.

\begin{figure} \centering
    \includegraphics[width=6cm,height=2.5cm,angle=0]{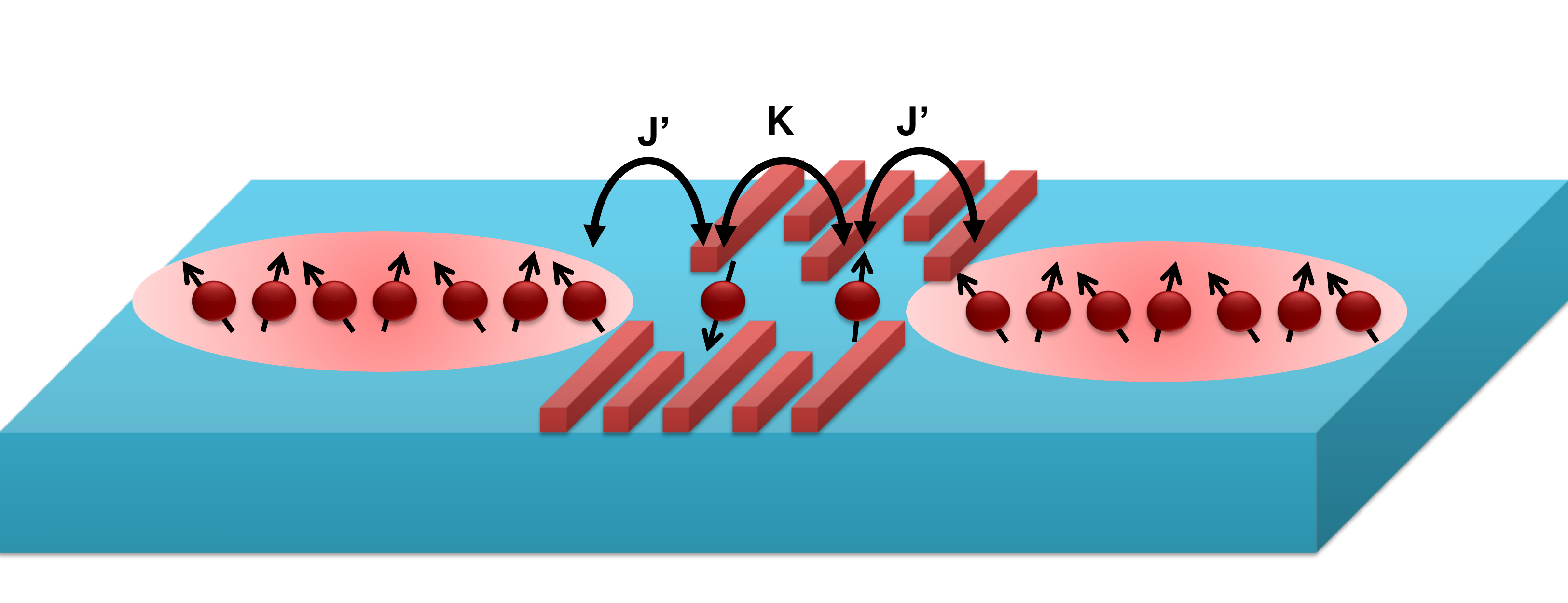}
    \caption{The two-impurity Kondo spin chain consists of two spin-1/2 impurities, each interacting with an array of spin-1/2 particles via a Kondo coupling $J'$. The inter-impurity RKKY coupling $K$, as well as the Kondo coupling $J'$, can be tuned via gates (shown as bars in the figure).}
     \label{fig1}
\end{figure}

\emph{Model.--} For the purpose of exploring quantum critical properties of the TIKM, it is sufficient to focus on the spin sector of the model. This can be emulated by the Kondo spin-chain Hamiltonian $H(K)= \sum_{m=L,R}H_m + H_I$ \cite{bayat-TIKM-12},  where
\begin{eqnarray} \label{Hamiltonian}
H_m &\!=\!&  J^{\prime} \left( J_1 \boldsymbol{\sigma}_1^m \!\cdot \!\boldsymbol{\sigma}_2^m + J_2 \boldsymbol{\sigma}_1^m \!\cdot \!\boldsymbol{\sigma}_3^m \right)+  \cr
&+&
J_1\sum_{i=2}^{N_m-1} \boldsymbol{\sigma}_i^m \!\cdot \!\boldsymbol{\sigma}_{i+1}^m+J_2\sum_{i=2}^{N_m-2} \boldsymbol{\sigma}_i^m \!\cdot \!\boldsymbol{\sigma}_{i+2}^m,
\cr
H_I &\!=\! & J_1K \boldsymbol{\sigma}_1^L \!\cdot \!\boldsymbol{\sigma}_1^R.
\end{eqnarray}
Here $m=L,R$ labels the left and right chains with $\boldsymbol{\sigma}_i^m$ the vector of Pauli matrices at site $i$ in chain $m$, and with $J_1$  ($J_2$)  nearest- (next-nearest-) neighbor couplings (see Fig.~\ref{fig1}).  In the following we set $J_1=1$ as our energy unit. The parameter $J'> 0$ plays the role of antiferromagnetic Kondo coupling and $K$ represents the dimensionless RKKY coupling between the impurity spins $\boldsymbol{\sigma}_1^L$ and $\boldsymbol{\sigma}_1^R$. The total size of the system is thus $N=N_L+N_R$. By fine tuning $J_2/J_1$ to the critical point $(J_2/J_1)_c=0.2412$ of the spin chain dimerization transition \cite{Okamoto,Eggert} all logarithmic scaling corrections vanish, allowing for an unambiguous fit of numerical data using the Density Matrix Renormalization Group (DMRG) \cite{white92,schollock05review,DMRGdechiara08}. Indeed, a DMRG study reveals that the Hamiltonian (\ref{Hamiltonian}) faithfully reproduces the features of the iQPT in the TIKM~\cite{bayat-TIKM-12}.

\emph{Thermodynamics of the quenched TIKM.--} We assume that the impurity coupling is initially $K$. The system is at zero temperature in its ground state $\ket{E_0(K)}$ with energy $E_0(K)$. The impurity coupling is then quenched
from $K$ to $K+\Delta K$, with the Hamiltonian $H(K)$ suddenly changing to $H(K+\Delta K)$. The work performed on the system becomes a stochastic variable $W$ described by the probability distribution function (PDF)~\cite{CampisiHTRMP11}
\begin{eqnarray}\label{E.WPDFSQ}
p(W)=\sum_{m} \left| \bra{E_m\rq{}} E_0(K) \rangle\right|^2 \delta\left[W-(E_m\rq{}-E_0(K))\right],
\end{eqnarray}
where  $\{ E_m\rq{}\}$ and $\{\ket{E_m\rq{}}\}$ are the eigenenergies and eigenvectors of $H(K+\Delta K)$, respectively.  Notice that the work PDF is an experimentally accessible quantity~\cite{WPDF,Roncaglia2014} and that from its knowledge all the statistical moments can be derived as $\average{W^n}=\int W^np(W) dW $.

Due to the nature of the sudden quench in the Hamiltonian, the system is driven out of equilibrium and, by means of the Jarzynski fluctuation-dissipation relation \cite{Jarzynski07}, it is possible to define the so-called {\textit{irreversible work}}:
\begin{equation}\label{E.Wirr}
 W_{irr}=\average{W}-\Delta F \geq0~,
\end{equation}
 where $\Delta F$ is the difference between the free energies after and before the quench. Since we assume zero temperature, $\Delta F$ is the difference of the post- and pre-quench ground state energies. The irreversible work has a simple physical explanation as the amount of energy which has to be taken out from the quenched system in order to bring it to its new equilibrium state which, for our case, is the ground state of $H(K+\Delta K)$~\cite{PlastinaetalPRL14}. For the instantaneous quantum quench we have
\begin{equation}\label{Wirr_general}
W_{irr}{=} \bra{E_0(K)}  H(K+\Delta K) \ket{E_0(K)}-E_0\rq{}(K+\Delta K)~,
\end{equation}
i.e., the irreversible work is given by the difference between the expectation value of the post-quenched Hamiltonian evaluated on the pre-quenched ground state and the post-quench ground state energy.



\begin{figure} \centering
    \includegraphics[width=9.5cm,height=7cm,angle=0]{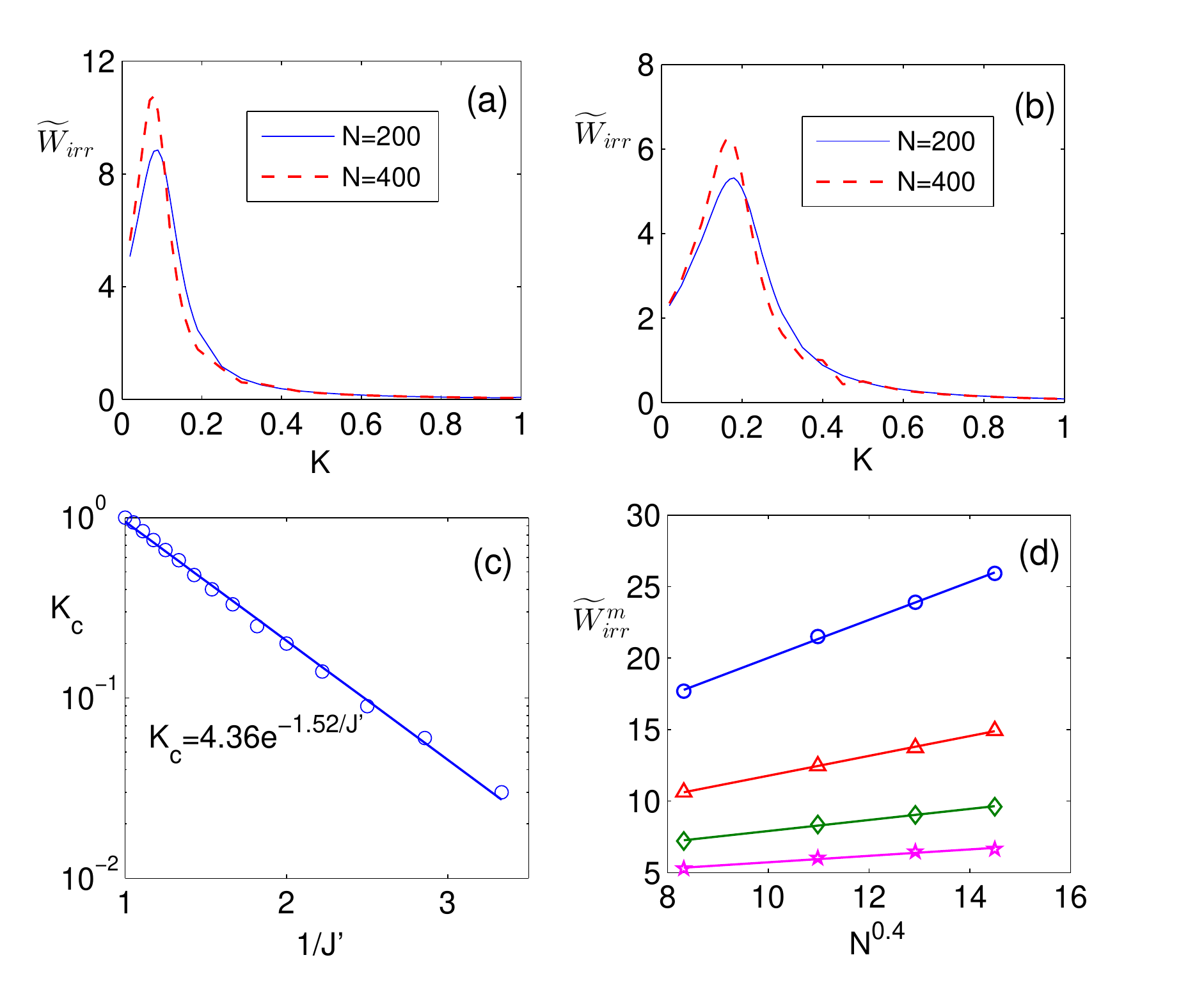}
    \caption{
The  irreversible work $\widetilde{W}_{irr}$ in terms of $K$ in a chain with (a) $J'=0.4$; (b) $J'=0.5$.
    (c) The critical coupling $K_c$ (blue circles) versus $1/J'$ in a semi-logarithmic scale and its exponential fit (blue line). (d) The maximum of the irreversible work $W_{irr}^m$ versus $N^{0.4}$ and the linear fits. From top to bottom:  $J'=0.4$; $J'=0.5$; $J'=0.6$ and $J'=0.7$.
    }
     \label{fig2}
\end{figure}

\emph{Scaling of the irreversible work.--} In order to capture the iQPT between the Kondo regime and the RKKY phase, we introduce the rescaled quantity $\widetilde{W}_{irr}=W_{irr}/\Delta K^2$
and study the variation of $\widetilde{W}_{irr}$ when the coupling $K$ is varied.
In this paper we only consider infinitesimal quantum quenches, $\Delta K \ll 1$. In Figs.~\ref{fig2}(a) and (b) we plot the irreversible work $\widetilde{W}_{irr}$ for two impurity couplings $J'=0.4$ and $J'=0.5$ respectively. It is clear from the plots that $\widetilde{W}_{irr}$ shows a sharp peak which becomes even more pronounced by increasing the system size $N$ (apart from slightly shifting towards lower values of $K$'s). This signifies that $\widetilde{W}_{irr}$ exhibits non-analytic behaviour at the critical point in the thermodynamic limit.  In finite-size systems, such as the ones considered here, the position of the peak determines the critical point $K_c$ which slowly moves towards the left by increasing $N$.

By considering the specific value of the RKKY coupling $K$ at which $\widetilde{W}_{irr}$ diverges, one can determine numerically the critical point $K_c$, which then shows a particular dependence on $1/J'$, just as the Kondo temperature $T_K$ (which sets the energy scale for the weak-to-strong of the renormalized Kondo coupling \cite{Jayaprakash81}).  This can be seen in Fig.~\ref{fig2}(c) in which the critical coupling $K_c$ is plotted as a function of $1/J'$. The linear curve in a semi-logarithmic scale confirms that $K_c\sim e^{-a/J'}\sim T_K$ for some constant $a$, in agreement with other studies of the two-impurity Kondo spin chain~\cite{bayat-TIKM-12,bayat-Schmidt-14}.

In the finite-size systems studied here, the divergence of $\widetilde{W}_{irr}$ at $K=K_c$ appears as a finite peak becoming more prominent for increasing system size, as shown in Figs.~\ref{fig2}(a) and (b). We define the maximum of the irreversible work as $\widetilde{W}_{irr}^m=\widetilde{W}_{irr}(K=K_c)$. Since $\widetilde{W}_{irr}^m$ increases by increasing the system size $N$ one can try to fit it by an algebraic map of the form
\begin{equation}\label{Wirr_max}
  \widetilde{W}_{irr}^m\sim N^{\lambda},
\end{equation}
where $\lambda$ is a positive exponent. In fact, a perfect match is found for various impurity couplings $J'$ by choosing $\lambda=0.4$ as depicted in Fig.~\ref{fig2}(d). Note that the
exponent $\lambda$ governs the scaling of a purely non-equilibrium quantity with system size. Note that, whereas for a {\textit{global}} quench the irreversible work is expected to have a functional dependence on the system size because in Eq.~(\ref{E.Wirr}) both the work and the free energy become extensive quantities, it is far from trivial that the same holds for a {\textit{local}} quench.  We explicitly focus on those impurity couplings for which the Kondo length is smaller than the system size $N$ in order to keep the Kondo physics valid in the Kondo regime (i.e. $K<K_c$). Nevertheless, in the TIKM here considered, the behavior of $ \widetilde{W}_{irr}^m$ is determined by the distinctive nature of the iQPT, where a local change in the RKKY coupling induces a global rearrangement of the ground state of the total system at criticality.

\begin{figure} \centering
    \includegraphics[width=9.5cm,height=4.5cm,angle=0]{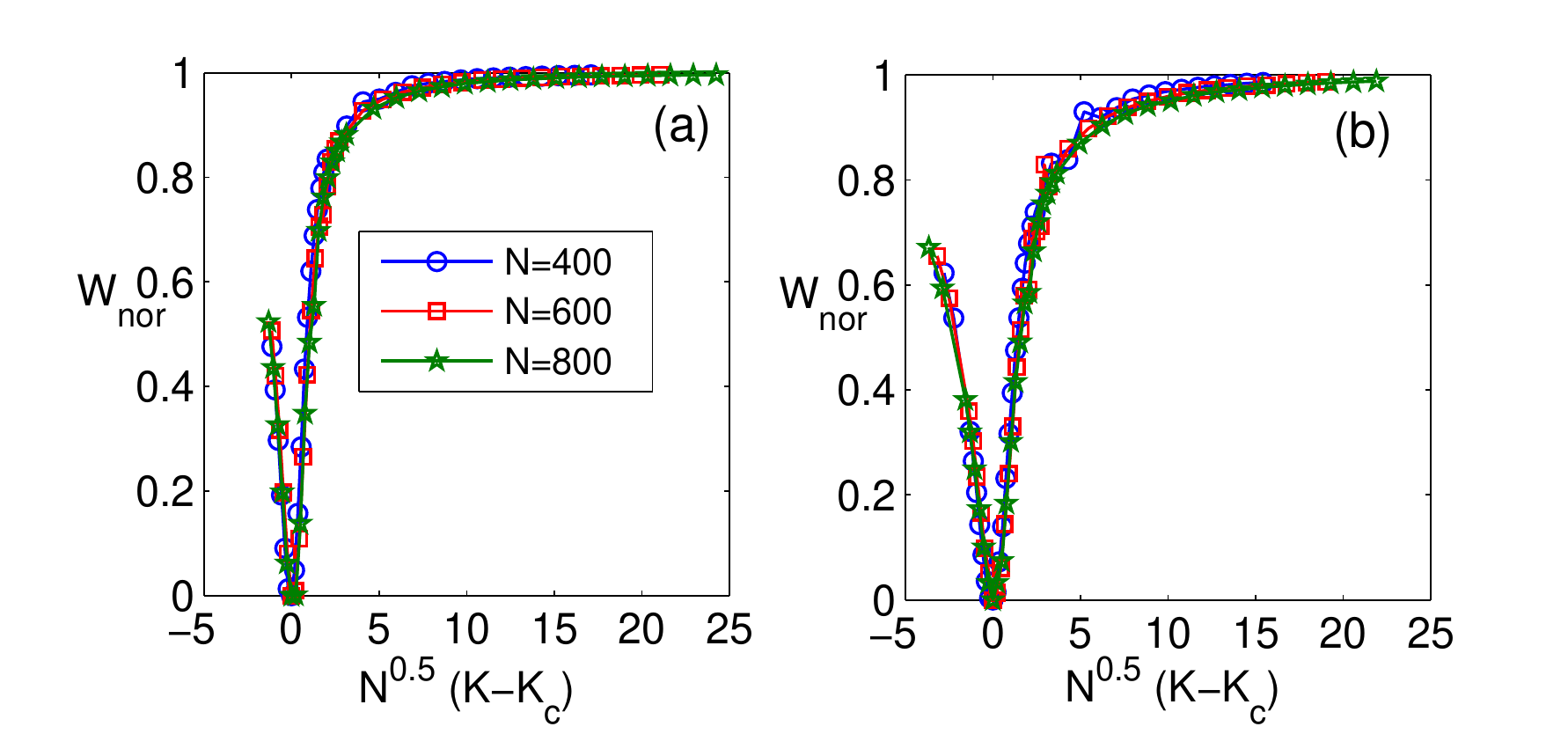}
    \caption{
Confirmation of the finite-size scaling of $W_{nor}$ for (a) $J'=0.4$; (b) $J'=0.5$. }
     \label{fig3}
\end{figure}

The above analysis  for $\widetilde{W}_{irr}$ suggests the ansatz:
\begin{equation}\label{Wirr_ansatz}
  \widetilde{W}_{irr}= \frac{A}{|K-K_c|^\kappa+B N^{-\lambda}},
\end{equation}
where $A$ and $B$ are two constants that may vary with $J'$. This ansatz is based on the fact that $\widetilde{W}_{irr}$ diverges in the thermodynamic limit as $\widetilde{W}_{irr}\sim |K-K_c|^{-\kappa}$, while for finite-size systems at $K=K_c$ it increases algebraically with the system size as in Eq.~(\ref{Wirr_max}).  In order to deal with the divergence more conveniently at the critical point we define a normalized function $W_{nor}= (\widetilde{W}_{irr}^m-\widetilde{W}_{irr})/\widetilde{W}_{irr}^m$.
Using the ansatz of Eq.~(\ref{Wirr_ansatz}) one can show that
\begin{equation}\label{Wnor_FS_scaling}
  W_{nor}=g(N^{\lambda/\kappa}|K-K_c|),
\end{equation}
where $g(x)$ is a scaling function which can be determined numerically. In order to evaluate the exponent $\kappa$ we search for the value of $\kappa$ such that the plots of $W_{nor}$ as a function of $N^{\lambda/\kappa}|K-K_c|$, for various system sizes $N$, collapse on each other, as shown in Figs.~\ref{fig3}(a) and (b) for two different impurity couplings $J'=0.4$ and $J'=0.5$ respectively. As is evident from the figure, using the predetermined exponent $\lambda=0.4$,
one finds that $\kappa=0.8$.

\begin{figure} \centering
    \includegraphics[width=9.5cm,height=7.5cm,angle=0]{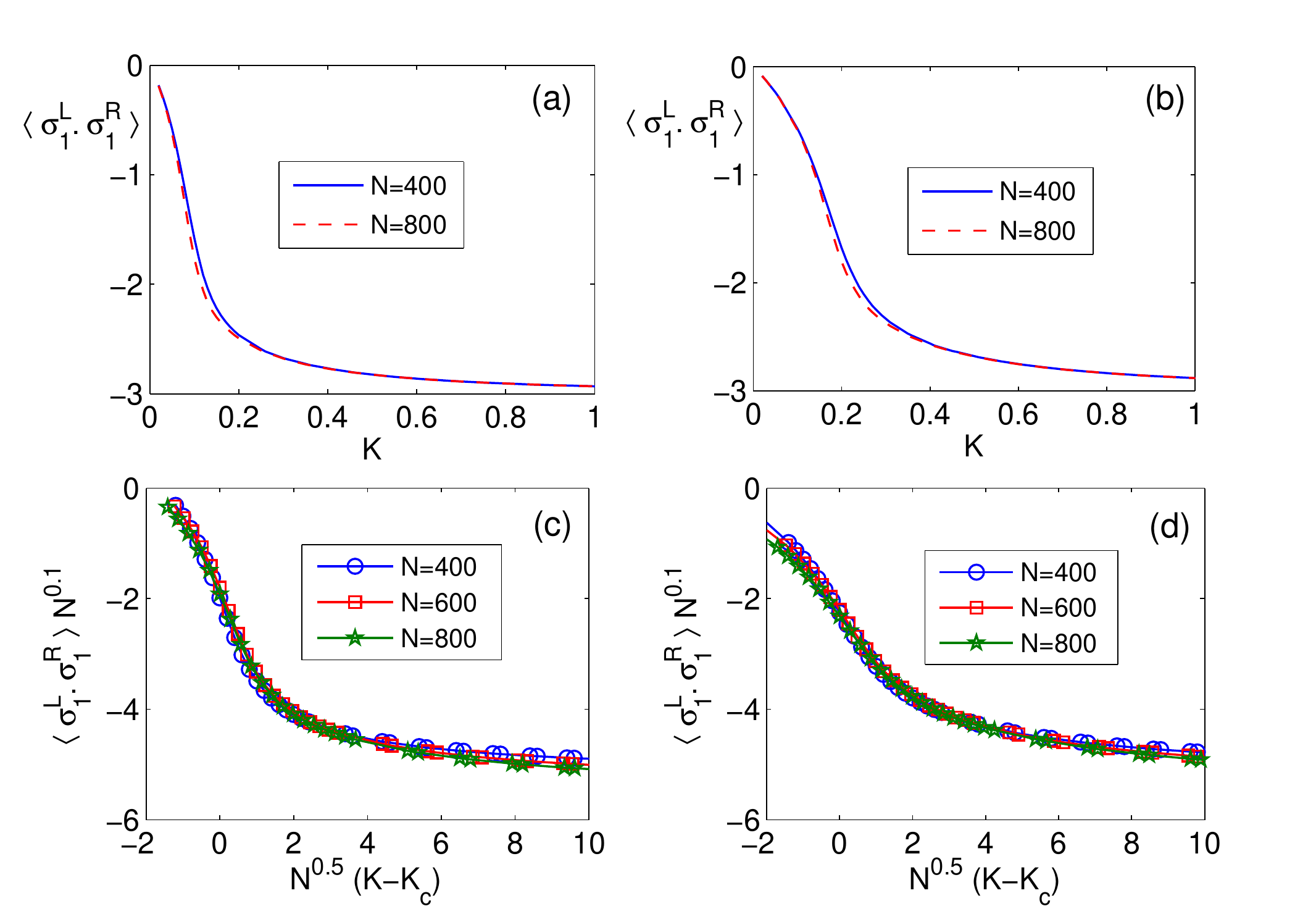}
    \caption{
Correlation function $\langle \boldsymbol{\sigma}_1^L\!\cdot\! \boldsymbol{\sigma}_1^R \rangle$ of the two impurities versus RKKY coupling $K$ in a chain with (a) $J'=0.4$; (b) $J'=0.5$.  The finite-size scaling for  $\langle \boldsymbol{\sigma}_1^L\!\cdot\! \boldsymbol{\sigma}_1^R \rangle$ with (c) $J'=0.4$; (d) $J'=0.5$.  }
     \label{fig4}
\end{figure}

The irreversible work has been measured in quantum mechanical setups using various methods \cite{Batalhao,An}.
Here we follow a different route, showing that,for small quenches, one can rely on measuring only the two-impurity correlation function $\average{\boldsymbol{\sigma}_1^L \!\cdot \!\boldsymbol{\sigma}_1^R}$ with respect to the ground state.  In a different study of the TIKM \cite{he2015}, a discontinuity of the two-impurity correlation function has been chosen to pinpoint the QPT point.

By expanding Eq.~(\ref{Wirr_general}) for small $\Delta K$ we obtain
\begin{equation}\label{Wirr_nor_Corr}
  \widetilde{W}_{irr}{=}{-}\frac{1}{2}\frac{\partial \average{\boldsymbol{\sigma}_1^L \!\cdot \!\boldsymbol{\sigma}_1^R}}{\partial K}.
\end{equation}
The divergence of $\widetilde{W}_{irr}$ at the critical point and using Eq.~(\ref{Wirr_nor_Corr}) suggests that the two-point impurity correlator $\langle \boldsymbol{\sigma}_1^L \!\cdot \!\boldsymbol{\sigma}_1^R \rangle$ mimics the behavior of an order parameter, capturing the quantum criticality and showing scaling behavior near the transition. In Figs.~\ref{fig4}(a) and (b) we plot the spin correlator $\langle \boldsymbol{\sigma}_1^L \!\cdot \!\boldsymbol{\sigma}_1^R \rangle$ versus the coupling $K$ for two impurity couplings $J'=0.4$ and $J'=0.5$ respectively. The correlator varies from $0$ (for $K=0$)  in the Kondo regime  to $\langle \boldsymbol{\sigma}_1^L \!\cdot \!\boldsymbol{\sigma}_1^R \rangle=-3$ (for very large $K$) deep in the RKKY phase.
To extract its scaling properties, we make the finite-size-scaling ansatz
\begin{equation}\label{Cor_FS_scaling}
  \langle \boldsymbol{\sigma}_1^L \!\cdot \!\boldsymbol{\sigma}_1^R \rangle = N^{-\beta/\nu}f(N^{1/\nu} |K-K_c|),
\end{equation}
where, in the limit $N\rightarrow \infty$, $\beta$ characterizes scaling of the correlator near criticality, $\langle \boldsymbol{\sigma}_1^L \!\cdot \!\boldsymbol{\sigma}_1^R \rangle \sim |K-K_c|^\beta$, $\nu$ is the exponent governing the divergence of the crossover scale $\xi \sim |K-K_c|^{-\nu}$~\cite{affleck1995,sela2011,Mitchell-sela2012}, and $f(x)$ is a scaling function. In order to determine these critical exponents we identify the values of $\beta$ and $\nu$ such that the plots of $\langle \boldsymbol{\sigma}_1^L \!\cdot \!\boldsymbol{\sigma}_1^R \rangle N^{\beta/\nu}$ as a function of $N^{1/\nu}|K-K_c|$ collapse to a single curve for arbitrary system sizes, as shown in
Figs.~\ref{fig4}(c) and (d). The best data collapse is achieved by choosing $\beta=0.2$ and $\nu=2$, which are in excellent agreement with the ones found from the Schmidt gap \cite{bayat-Schmidt-14}.

Furthermore, as an alternative way of computing the scaling of the irreversible work  $\widetilde{W}_{irr}$, one may directly differentiate both sides
of Eq.~(\ref{Cor_FS_scaling}) with respect to the RKKY coupling $K$ to get
\begin{equation}\label{Wirr_FS_scaling}
  \widetilde{W}_{irr} \sim \partial_K \langle \boldsymbol{\sigma}_1^L \!\cdot \!\boldsymbol{\sigma}_1^R \rangle \sim N^{(1-\beta)/\nu}f'(N^{1/\nu} |K-K_c|),
\end{equation}
where $f'(x)=df/dx$. The finite-size scaling of Eq.~(\ref{Wirr_FS_scaling}) implies that $\widetilde{W}_{irr} \sim |K-K_c|^{\beta-1}$, which then leads to
\begin{equation}\label{exponent1}
  \kappa=1-\beta.
\end{equation}
Moreover, comparing Eq.~(\ref{Wirr_FS_scaling}) and Eq.~(\ref{Wnor_FS_scaling}), we obtain another constraint  between the exponents as
\begin{equation}\label{exponent2}
  \kappa=\lambda \nu.
\end{equation}
Eqs.~(\ref{exponent1}) and (\ref{exponent2}) are indeed satisfied for the values found in our numerical analysis as $\lambda=0.4$, $\nu=2$, $\beta=0.2$ and $\kappa=0.8$, confirming our scaling ans\"atze.

It is worth emphasizing that in our local quench problem the energy change, for every finite quench, is always finite and, for an infinitesimal quench $\Delta K$,  the irreversible work can be approximated by $W_{irr} \simeq - \Delta K \Delta \average{\boldsymbol{\sigma}_1^L \!\cdot \!\boldsymbol{\sigma}_1^R}\!/2$. Since $\average{\boldsymbol{\sigma}_1^L \!\cdot \!\boldsymbol{\sigma}_1^R}$ varies between $0$ and $-3$, then $W_{irr}\leq -3\Delta K/2$, which vanishes for $\Delta K\to 0$.  As a consequence, the un-rescaled irreversible work $W_{irr}$ shows no divergences even as $N\rightarrow \infty$.

\emph{Variance of work.--} The variance of work defined as $\Delta W^2=\average{W^2}-\average{W}^2$ is another important non-equilibrium quantity.
 For convenience,  we also rescale the variance as $\Delta \widetilde{W}^2{=}\Delta W^2/\Delta K^2$. For a sudden quench one can show that
 $\Delta \widetilde{W}^2{=}3{-}2\average{\boldsymbol{\sigma}_1^L \!\cdot \!\boldsymbol{\sigma}_1^R}{-}\average{\boldsymbol{\sigma}_1^L \!\cdot \!\boldsymbol{\sigma}_1^R}^2$.
The derivative of the rescaled variance with respect to $K$ becomes
$\partial_K (\Delta \widetilde{W}^2) = 4 \left(1+ \average{\boldsymbol{\sigma}_1^L \!\cdot \!\boldsymbol{\sigma}_1^R}\right) \widetilde{W}_{irr}$.
Since the correlation function $\average{\boldsymbol{\sigma}_1^L \!\cdot \!\boldsymbol{\sigma}_1^R}$ is always finite, both $\widetilde{W}_{irr}$ and   $\partial_K (\Delta \widetilde{W}^2)$ diverge at the critical point in the thermodynamic limit. Moreover, $\Delta \widetilde{W}^2$ takes its maximum slightly before $K_c$ where $\average{\boldsymbol{\sigma}_1^L \!\cdot \!\boldsymbol{\sigma}_1^R}=-1$ and the two impurities start to get entangled.


The irreversible work and the variance of work encode the dynamical response of an equilibrium ground state to perturbations. Due to the sudden quench approximation, both quantities are given by expectation values over ground states.  Moreover, for small quenches, one can rely on first-order perturbation theory, and, similar to linear response theory, the irreversible work is given by the susceptibility of the ground state two-impurity correlation function. Nevertheless, we should also point out that whereas a temperature can be associated to the initial state ($T = 0$ in our analysis), the same does not hold after the quench has been performed.

\emph{Conclusion.--} In this letter, we have shown that both irreversible work and work variance, as non-equilibrium quantities, signal the impurity quantum phase transition between the Kondo and RKKY regimes in the TIKM. Both quantities 
exhibit scaling at the quantum critical point, and their corresponding critical exponents have been determined.
Importantly, all out-of-equilibrium quantities considered are amenable to experimental observation in solid-state nanostructures or ultra cold atoms, since ultimately it is sufficient to measure a two-point  spin correlation function.

\emph{Acknowledgements.--} This work is supported by the John Templeton Foundation (grant ID 43467), the EU Collaborative Project TherMiQ (Grant Agreement 618074), and the Swedish Research Council (Grant No. 621-2014-5972). The research leading to these results has received funding
from the European Research Council under the European
Union\rq{}s Seventh Framework Programme (FP/ 2007-2013)/ERC Grant Agreement n. 308253.  SP is supported by a Rita Levi-Montalcini fellowship of MIUR. AB and SB acknowledge the EPSRC grant $EP/K004077/1$. PS thanks the
Ministry of Science, Technology and Innovation of Brazil,
MCTI and UFRN/MEC for financial support and CNPq for
granting a \lq{}\lq{}Bolsa de Produtividade em Pesquisa\rq{}\rq{}. All authors thank the International Institute of Physics in Natal, Brazil, where this work has been completed.

\end{document}